\documentclass[final,a4paper]{article}
\pdfoutput=1 

\usepackage[T1]{fontenc} 

\usepackage{setspace}

\usepackage[color]{showkeys}
\usepackage{graphicx}
\usepackage{physics}
\usepackage{dsfont}
\usepackage{epsfig,subfigure,float}
\usepackage{epstopdf}
\usepackage[normalem]{ulem}

\usepackage{jheppub} 
                     
\definecolor{refkey}{gray}{0.75}
\definecolor{labelkey}{RGB}{155,48,48}
\renewcommand*\showkeyslabelformat[1]{%
  \fbox{\parbox[t]{0.8\marginparwidth}{\raggedright\normalfont\scriptsize\url{#1}}}}

\def\beq{\begin{eqnarray}}\def\eeq{\end{eqnarray}}
\def\be{\begin{equation}}\def\ee{\end{equation}}

\def\be{\beta}
\def\de{\delta}
\def\ep{\epsilon}

\def\Ga{\Gamma}

\def\si{\sigma}

\title{A Note on S-Matrix Bootstrap for Amplitudes with Linear Spectrum}

 \author{Pranjal Nayak,}
 \author{Rohan R. Poojary,}
 \author{Ronak M Soni}
 \affiliation{\it Department of Theoretical Physics,
 Tata Institute of Fundamental Research,\\  Colaba, Mumbai, 400005, India}

\vspace{5mm}

\vspace{1cm}

\emailAdd{pranjal@theory.tifr.res.in}
\emailAdd{ronp@theory.tifr.res.in}
\emailAdd{ronak@theory.tifr.res.in}

\abstract{
  We work out constraints imposed by channel duality and analyticity on tree-level amplitudes of four identical real scalars, with the assumptions of a linear spectrum of exchanged particles and Regge asymptotic behaviour.
  We reduce the requirement of channel duality to a countably infinite set of equations in the general case. We show that channel duality uniquely fixes the soft Regge behaviour of the amplitudes to that found in String theory, $(-s)^{2t}$.
  Specialising to the case of tachyonic external particles, we use channel duality to show that the amplitude can be any one in an infinite-dimensional parameter space, and present evidence that unitarity doesn't significantly reduce the dimension of the space of amplitudes.
}

\begin{document}
\maketitle
\flushbottom

\section{Introduction}
String theory arose from an attempt to write down scattering amplitudes for string interactions from consistency conditions rather than from a Lagrangian, also known as the S-Matrix bootstrap program.
This program, however, was eventually abandoned in favour of $SU(3)$ Yang-Mills theory.
The amplitudes written by Veneziano \cite{Veneziano1968} and generalised by Virasoro \cite{Virasoro1:PhysRev.177.2309,Virasoro2:PhysRevD.1.2933}, while they didn't prove very useful for understanding strong interactions, eventually gave rose to string theory, which then shed its roots in this program to become a field in its own right.

This program has, in some sense, seen a revival in recent years, both indirectly through applications in conformal bootstrap \cite{Paulos:2016fap,Paulos:2016but,Li:2017lmh} and more directly through a striking result about the three-point functions between two gravitons and higher-spin fields: \cite{Camanho:2014apa} proved that the three-point functions must either match Einstein gravity, or there must exist an infinite tower of higher-spin fields in the theory.

This last result raises a very interesting question.
This question is predicated on the fact that there are only a few classical, tree-level, amplitudes known that have an infinite tower of higher spins: as many as there are different string theories.
While, quantum-mechanically, string theory is plagued by a large number of possible compactifications -- the so-called ``landscape'' problem --, each string theory gives a unique tree-level amplitude, for the simple reason that without the moduli space integrations required at loop level the different dimensions of space-time correspond to decoupled CFTs on the worldsheet.\footnote{We thank Shiraz Minwalla for explaining this fact, and consequently suggesting this problem, to us.}
The question is this: given that there are so few examples known of tree-level graviton amplitudes, could it be that these are the only examples?
In other words, could it be that the string theories are the only consistent extensions of classical gravity?

The obvious question is: consistent with what?
A minimal list of conditions would include: Lorentz-invariance, causality, unitarity and crossing symmetry.
So, we could ask the question of what the most general four-point graviton scattering amplitude consistent with this minimal set of consistency conditions is.

In this note, however, we try to head towards this problem via a simpler, more restricted problem.
We consider the scattering of four identical scalars (instead of gravitons), and further assume a linear spectrum of exchanged particles and Regge behaviour at large energies.
We come to the conclusion that crossing symmetry restricts the Regge asymptotic behaviour to be $A(s,t) \xrightarrow{s \to \infty} (-s)^{2t}$ but still allows for an infinite-dimensional parameter space of amplitudes, and argue numerically that unitarity doesn't significantly reduce the dimensionality of this allowed space.

This problem has been addressed recently, using very different methods, in \cite{Caron-Huot:2016icg,Cardona:2016ymb, Sever:2017ylk,boels2014string}.
The results of \cite{Caron-Huot:2016icg,Cardona:2016ymb} are more or less assumed in our work, in the assumption of linear spectrum.
Those of \cite{Sever:2017ylk} are not relevant for this work, since the case of linear spectrum is a very degenerate one and that work goes beyond this case.
Finally, the authors of \cite{Boels:2014dka} are able to show that string theory amplitudes can be derived from monodromy and BCFW recursion relations. It would be interesting to use the recursion relations that they have developed for the string theory amplitudes to constraint the higher-point scattering amplitudes using our techniques.

\subsection{Organisation and Summary of the Note}
More precisely, for tree-level four-point amplitudes of four identical scalars, we impose the conditions that its behaviour at large $s$ is\footnote{This equation is missing some factors. See \eqref{eqn:regge-bahviour} for a more precise statement.}
\begin{equation}
  A(s,t) \xrightarrow{s \to \infty} (-s)^{- k(-t)},
  \label{eqn:intro-regge}
\end{equation}
that the mass-squareds of exchanged particles are evenly spaced,
\begin{equation}
  m_{n}^{2} = \frac{n - \alpha_{0}}{\alpha'},
  \label{eqn:intro-spectrum}
\end{equation}
and some other conditions listed in section \ref{sec:postulates}, and investigate the restrictions imposed first by crossing symmetry and then by unitarity.

It should be mentioned here that, throughout this note, we work not with the standard Mandelstam variables $s,t,u$ but with shifted and rescaled versions of them $a,b,c$ (see \eqref{eqn:a-defn}) such that the poles from the intermediate particles going on-shell are at $a = 0,-1,-2,\cdots$ for the $s$-channel, $b=0,-1,-2\cdots$ for the $t$-channel and $c=0,-1,-2\cdots$ for the $u$-channel.
For clarity, however, we write the remainder of this introduction in terms of the standard Mandelstam variables while being cavalier about factors.

\begin{figure}[H]
  \centering
  \includegraphics{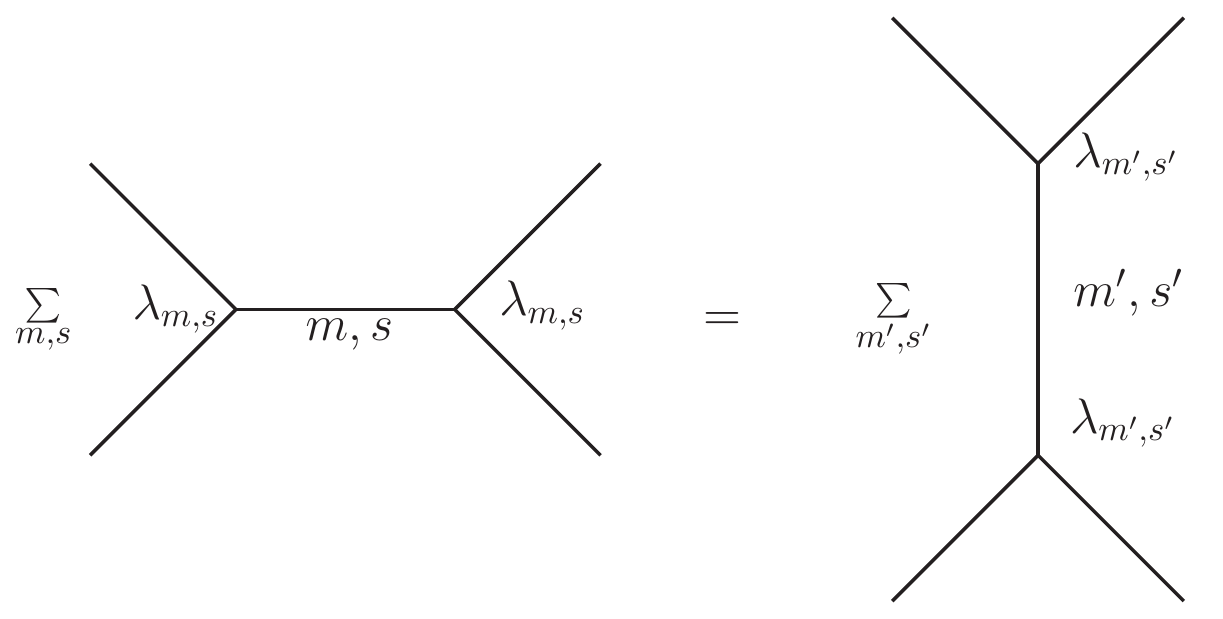}
  \caption{Crossing symmetry in theories with Regge asymptotic behvaiour is imposed by ``channel duality,'' the requirement that the sum over $s$-channel poles be equal to the sum over $t$-channel poles.}
  \label{fig:crossing}
\end{figure}

With the Regge behaviour condition, it is known that crossing symmetry must be implemented not by an independent sums of $s$-channel, $t$-channel and $u$-channel diagrams but by ``channel duality:'' the $t$-channel poles have to be hidden in an infinite sum over $s$-channel poles, as illustrated in figure \ref{fig:crossing}.
This is because a $t$-channel diagram with an exchange of a particle of spin $l$ behaves at large $s$ as $s^{l}$, which necessarily overpowers the exponential falloff that is the Regge behaviour, and therefore it must be possible to write the entire amplitude in the region $s>0$ without summing over $t$-channel diagrams.
See  \cite{Caron-Huot:2016icg,Cardona:2016ymb} for further discussion.

In section \ref{sec:channel-duality}, we reduce the condition of channel duality to a countably infinite set of equations \eqref{eqn:gj-conds} in terms of the values at a discrete set of points of the coefficients \eqref{eqn:1byn} of the Laurent expansion of the amplitude about $a = \infty$ (they are still functions of $t$).
We also obtain a physical interpretation of the function $k(-t)$ that appears in the exponent in the Regge behaviour, that $-k(-n)$ is the maximum spin exchanged at level $n$ (where the lightest exchanged particle corresponds to $n=0$).
However, we aren't able to proceed beyond this at this level of generality.

To be able to proceed, we impose the further condition that the function $k$ is linear in its argument in section \ref{sec:lin-asymptotics}.
First, using various complex analysis techniques (mostly Carlson's theorem), we're able to show that the function $k$ can't be just any linear function but has to be $k(-t) = 2 (-t) + l$, which is exactly the sort of behaviour shown by the amplitudes in string theory!
This is one of the main results of this note.\footnote{This result was indicated by, though not quite proved, by previous work \cite{Matsuda:PhysRev.185.1811}.}
The final conclusion of section \ref{sec:lin-asymptotics}, the channel duality equations \eqref{eqn:defns}, \eqref{eqn:rmes} and \eqref{eqn:spes} are among the other major results of this note; they are necessary and sufficient conditions for an amplitude that satisfies our assumptions to be channel-dual.
In the discussion subsection, we try to relate each of these equations to some physical meaning.

In section \ref{sec:lol-max}, we turn to solving these equations.
While these equations are rather hard to solve, we show that given any channel-dual amplitude $A_{a,b,c}$ there is an infinite-dimensional parameter space of amplitudes $\sum_{m=0}^{\infty} a_{m} A(a+m,b+m,c+m)$ that are also all channel dual with the same poles.
While this is obvious, we also show it using our bootstrap equations, for no other reason than that the proof is simple and pretty. That these class of 4-point amplitudes satisfy all the assumptions of dual amplitudes is known from the works of \cite{Khuri:1970ax,Weimar:1974pi,COON1969669,Matsuda:PhysRev.185.1811}.

Finally, in section \ref{sec:pictures} we turn to the question of unitarity and show in the case of the Virasoro-Shapiro amplitude and the dilaton amplitude in closed bosonic string theory that many perturbations of these base amplitudes seem to be consistent with unitarity.
This section consists only of numerical arguments.

Appendix \ref{sec:kinematics} summarises some useful facts about the kinematics of the amplitudes we consider here.
Appendix \ref{sec:explicitVS} shows explicitly how our analytic continuation and other techniques and results apply to the Euler beta function, which is the building block of the Veneziano amplitude.

Mathematica files containing parts of relevant computations of this paper are available as ancillary files on arXiv.

\section{Postulates} \label{sec:postulates}
We begin by laying out the properties that we require the amplitude to satisfy.
The assumptions broadly fall into three categories: legitimate choices about the problem to consider, physically necessary conditions, and conditions that aren't required but give us a lot of control of the problem.
We indicate while listing which type each assumption is in, except where obvious.

The physical situation under consideration is the 2-2 tree-level scattering of four identical scalars of mass $M_{ext}$ in a $D$-dimensional Minkowski spacetime.
We number the incoming particles $1$ through $4$, with $k_i$ labelling their respective momenta. We take all four momenta to be ingoing and the momentum conservation with this convention reads, $k_1+k_2+k_3+k_4=0$.
We take the metric to have mostly positive signature.
The Mandelstam variables are
\begin{align}
  s &=- (k_{1} + k_{2})^{2} \nonumber\\
  t &= - (k_{1} + k_{3})^{2} \nonumber\\
  u &= - (k_{1} + k_{4})^{2}.
  \label{eqn:stu-defn}
\end{align}

Having set the stage, let us list the assumptions.
For a more detailed review, see \cite{Veneziano1974199}.
\begin{enumerate}
  \item \textbf{Lorentz-Invariance}: The amplitude $A$ is only a function of the Mandelstam variables $s,t,u$, and momentum is conserved at every vertex.
  \item \textbf{Causality, or analyticity}: In the two-dimensional complex $s,t$ plane the singularities occur only when one or more intermediate particles go on-shell.
    This allows both tree-level and loop diagrams (in which case the singularities are branch cuts).
    In particular, the singularities in the amplitude appear only for real values of $s,t$ or $u = 4M_{ext}^{2} - s- t$.
  \item \textbf{Restriction to Tree-Level Amplitudes}: The amplitude is a sum of only tree-level diagrams.
    Combined with the previous assumption, this means that the only singularities of the amplitude are poles, at values of $s,t,u$ equal to the mass-squared of a particle in the spectrum of the theory.
  \item \textbf{Unitarity, or Cutting Rules}: Unitarity of scattering amplitudes is generally ensured by cutting rules.
    Suppose there are particles of spin $0$ through $L$ at some mass $m$.
    Then, the residue of the amplitude at the pole $s = m^{2}$ must be
    \begin{equation}
      Res_{s=m^{2}} A(s,t) = \sum_{l=0}^{L} \lambda_{m,l}^{2} C_{l}^{\left( \frac{D-3}{2} \right)} \left(1 + \frac{2t}{s-M_{ext}^{2}}\right),\quad \lambda_{m,l}^{2} \ge 0.
      \label{eqn:res-gen-form}
    \end{equation}
    Here, each $\lambda_{m,l}^{2}=\sum_{i} \lambda_{m,l,i}^{2}$ is the sum of squares of cubic couplings $\lambda_{m,l,i}$ of two external scalars and all the particles of mass $m$ and spin $l$ (which we have labelled by $i$), and the functions $C_{l}^{(\alpha)}$ are Gegenbauer polynomials (which reduce to Legendre polynomials for $D=4$); the argument of the Gegenbauers is $\cos \theta$, where $\theta$ is the scattering angle in the centre-of-momentum frame, see appendix \ref{sec:kinematics} for details.

    It should be noted that this requirement is only a necessary and not a sufficient condition for unitarity, because of the possibility of many particles with the same mass and spin.
    In the case when there are multiple such particles, it doesn't restrict all the $\lambda_{m,l,i}^{2}$s to be positive but only their sum, $\sum_{i} \lambda_{m,l,i}^{2}$.
  \item \textbf{Crossing Symmetry}: The amplitude should be invariant under exchange of any pair of external particles, since all four particles are identical.
    In terms of the Mandelstam variables, this means that the amplitude should, as a function, satisfy the relations
    \begin{equation}
      A(s,t) = A(t,s) = A(s,u = -4M_{ext}^{2} - s - t).
      \label{eqn:crossing}
    \end{equation}
  \item \textbf{Linear Spectrum}: The mass-squareds of the exchanged particles are spaced linearly, that is
    \begin{equation}
      m_{n}^{2} = \frac{n-\alpha(0)}{\alpha'}, \quad \big/ n \in \{0,1,2\cdots\}.
      \label{eqn:masses}
    \end{equation}
    This is the first really non-trivial assumption here, the first in the class of ``conditions that aren't required but give a lot of control over the problem;'' the ones earlier were all properties that we must require of all tree-level amplitudes.

    To take advantage of this simple behaviour, we define proxy variables $a,b,c$\footnote{These were called $-\alpha(s),-\alpha(t),-\alpha(u)$ in the old bootstrap literature, but we use this notation to avoid clutter, and also because these are the natural variables that turn up in the Virasoro-Shapiro amplitude, as named in \cite{polchinski1998string}. } for $s,t,u$ as
    \begin{equation}
      a = - \alpha' s - \alpha(0),
      \label{eqn:a-defn}
    \end{equation}
    and similarly for $b$ and $c$, so that the poles in the amplitude are at $a=-n$,
    \begin{equation}
      s = m_{n}^{2} \quad \Leftrightarrow \quad a = -n
      \label{eqn:a-defining-property}
    \end{equation}
    and similarly for $t,u$ and $b,c$.
    Also note that this $\alpha'$ isn't the constant that appears in the string action, but just the inverse of the level spacing; in particular for closed bosonic string theory our $\alpha'$ is related to that one as $\alpha'_{us} = \frac{1}{4}\alpha'_{closed\ bosonic}$.

    Because $s+t+u = 4M_{ext}^{2}$ is a constant, so is $a+b+c$.
    We call this constant $P$ for the remainder of this note,
    \begin{equation}
      a+b+c \equiv P = - 4 \alpha' M_{ext}^{2} - 3 \alpha(0).
      \label{eqn:P-defn}
    \end{equation}

    For later convenience, we note that in these variables, the physical s-channel scattering region is given by $a<0, -\alpha(0) < b < -a - \alpha(0)$.
  \item \textbf{Regge Asymptotic Behaviour, or ``Analyticity of the Second Kind''}: At large $s$ (or $t$ or $u$), the amplitude behaves as
    \begin{equation}
      A(a,b,c) \xrightarrow{a \to -\infty} a^{-k(b)} \sim (-s)^{-k(-t)}.
      \label{eqn:regge-bahviour}
    \end{equation}
    where $k(b) > 0$ in the physical s-channel region, $b>0$.
    
    This appears to be non-analytic at $a = - \infty$ because $k(b)$ need not be an integer. However, this is not a true non-analyticity of  the amplitude, but the result of the fact that we are restricting ourselves to physical scattering wedge in writing the above asymptotic behaviour.
    
    In particular, we will heavily use that the amplitude admits a Laurent expansion around the infinity once we factor out this apparent non-analyticity. Before we go ahead to discuss the implications of this assumption, we wish to mention that ``analyticity of the second kind'' often refers to a class of assumptions on the asymptotic behaviour of the amplitudes. These assumptions are used as additional postulates that differentiate the amplitudes of weakly interacting theories like QED and the weak force from those of strongly interacting theories, \cite{Veneziano1974199}. Moreover, these assumptions are independent of the bounds that are implied on the asymptotic behaviour of any general amplitude that obeys all the previous assumptions, like the Froissart and Martic bounds.

    The above assumption, \eqref{eqn:regge-bahviour}, is inconsistent with the amplitude being a sum of separate diagrams in the $s,t$ and $u$ channels, because a $t$-channel diagram of spin $l$ behaves as $s^{l}$ at large $s$, and the fact that $l \ge 0$ means that this will overpower the Regge falloff; see \cite{Caron-Huot:2016icg,Cardona:2016ymb} for more details.
    So, the assumption of Regge behaviour means that crossing symmetry is implemented by \textbf{channel duality}; the sum over all $s$-channel diagrams has all the $t$-channel poles hidden in it.
    The main thrust of this paper is to understand how they're hidden in it.
\end{enumerate}

Before going ahead, we note that these assumptions are all true for the Virasoro-Shapiro amplitude.
The Virasoro-Shapiro amplitude, which is the scattering amplitude of four tachyons in bosonic string theory, is
\begin{equation}
  \frac{\Gamma(a) \Gamma(b) \Gamma(c)}{\Gamma(a+b) \Gamma(b+c) \Gamma(c+a)},
  \label{eqn:vs}
\end{equation}
with
\begin{equation}
  a = - \frac{\alpha'}{4} s - 1,
  \label{eqn:a-vs}
\end{equation}
where $\alpha'$ is not the Regge slope from \eqref{eqn:a-defn} but is related to the inverse of string tension.
In this case, the external mass-squared is $M_{ext}^{2} = - \frac{4}{\alpha'}$ and $a+b+c = 1$.
The asymptotic behaviour is
\begin{equation}
  A(a,b,c) \xrightarrow{a \to \infty} a^{-2b},
  \label{eqn:vs-regge}
\end{equation}
as can be easily shown using Stirling's approximation.

\section{Channel Duality Equations} \label{sec:channel-duality}
Having set up the problem, we now try to understand the constraints imposed by channel duality.
In this section, we show that the constraints can be reduced to a countably infinite set of equations that have to be simultaneously satisfied.
The strategy will be to take the pole-sum form in the $a$ and $c$ channels,
\begin{equation}
  A(a,b) = \sum_{n=0}^{\infty} \frac{f_{n} (b)}{a+n} + \frac{f_{n} (b)}{c+n},\quad \Re b > 0,
  \label{eqn:pole-sum-form}
\end{equation}
and recreate the poles in the $b$-channel by a suitable analytic continuation.
Since there are no explicit poles in $b$,\footnote{Technically, the $1/(P-a-b+n)$ part has singularities in the $b$ complex plane, but we're only looking for $b$-channel poles, whose positions are independent of $a$} it must be the case that these poles come from the infinite sum.
To see that this is the case, we can ask how the above function can diverge for a particular value of $b$; the answer is clearly that the infinite sum may diverge for particular values of $b$.
In particular, that means that this sum must converge for all positive values of $b$, where there are no poles (we are again ignoring the $c$-channel poles here, which are aleady explicit in the above expression).
In \autoref{sec:explicitVS}, we demonstrate the ideas discussed in this and the next section by implementing them on Euler beta functions, which are the building blocks of Veneziano amplitudes.

To make this explicit, we expand the residue in a Laurent series around $n = \infty$,
\begin{equation}
  f_{n} (b) = \sum_{j=0}^{\infty} g_{j} (b) n^{-k(b) - j}.
  \label{eqn:1byn}
\end{equation}
To see that this is the correct leading behaviour, we note that the asymmptotic behaviour in $a$ is innherited by $n$, since at $a = -n - \varepsilon$, $\varepsilon < 1$, the amplitude can be approximated by the closest term in the pole-sum
\begin{equation}
  A(a,b) \approx \frac{1}{\varepsilon} f_{n} (b) = \frac{1}{\varepsilon} f_{a+\varepsilon}(b),
  \label{eqn:asymptotic-translation}
\end{equation}
so that this reproduces the Regge behaviour we have assumed.
This expansion has to be true, because all dependence of the amplitude \eqref{eqn:pole-sum-form} that isn't just $a^{-1}$ can only come from here. 
Also note that we chose to do the expansion in $n$ instead of $-n \sim a$; this is a matter of convenience and will have implications later.

We proceed by subsituting the $1/n$ expansion of the residue \eqref{eqn:1byn} into the pole-sum form of the amplitude \eqref{eqn:pole-sum-form},
\begin{equation}
  A(a,b) \!=\! \sum_{n=2}^{\infty} \!\! \left( \sum_{j=0}^{\infty} g_{j} (b) n^{-k(b)-j} \!\! \right) \left( \frac{1}{a+n} + \frac{1}{P-a-b+n} \right) + \sum_{n=0}^{1} f_{n} (b) \left( \frac{1}{a+n} + \frac{1}{P-a-b+n} \right),
  \label{eqn:1byn-expanded-1}
\end{equation}
where the $n \le 1$ terms have been split off because they're clearly outside the radius of convergence of the $1/n$ expansion.
However, since the split-off part is clearly regular in $b$ -- it's just a polynomial --, we  may safely ignore them; to put this another way, the divergence must come from the tail end of the sum so dropping a finite sum in the beginning should not be a problem.

For all $n > |a|$, we can also expand
\begin{equation}
  \frac{1}{a+n} = \sum_{r=0}^{\infty} \frac{(-a)^{r}}{n^{r+1}},
  \label{eqn:1byn-for-pole}
\end{equation}
and similarly for the $1/(c+n)$ term.

Plugging this into the amplitude \eqref{eqn:1byn-expanded-1} gives\footnote{Here $[a]$ denotes smallest integer greater than $a$.}
\begin{align}
  A(a,b) &= \sum_{n=[a]}^{\infty} \left\{ \left( \sum_{j=0}^{\infty} g_{j} (b) n^{-k(b)-j} \right) \sum_{r=0}^{\infty} \frac{(-a)^{r}}{n^{r+1}} \right\} \nonumber \\
  & \qquad + \sum_{n=[P-a-b]}^{\infty} \left\{ \left( \sum_{j=0}^{\infty} g_{j} (b) n^{-k(b)-j} \right) \sum_{r=0}^{\infty} \frac{(a+b-P)^{r}}{n^{r+1}} \right\} + reg. \nonumber\\
  &= \sum_{n=1}^{\infty} \left( \sum_{j=0}^{\infty} g_{j} (b) n^{-k(b)-j} \right) \left( \sum_{r=0}^{\infty} \frac{(-a)^{r} + (a+b-P)^{r}}{n^{r+1}} \right) + reg.
  \label{eqn:1byn-expanded-2}
\end{align}
where we have again ignored all finite sums in $n$.
We can rearrange the sums in the last expression to get
\begin{equation}
  A(a,b) = \sum_{j,r=0}^{\infty} g_{j} (b) \left\{ (-a)^{r} + (a+b-P)^{r} \right\} \sum_{n=1}^{\infty} n^{-k(b) - j - r - 1} + reg.
  \label{eqn:1byn-expanded-3}
\end{equation}

So far, all our manipulations have only been valid for $\Re b > 0$.
To find the poles in $b$, which resisde at $b = -n$, we need to analytically continue the expression \eqref{eqn:1byn-expanded-3} to the left half plane of $b$.
The innermost sum in \eqref{eqn:1byn-expanded-3} diverges for all $b$ such that $k(b) \le 0$; this is an artefact of the fact that the expression isn't valid in that region.
To analytically continue, we replace the sums over $n$ by Riemann zeta functions -- that is, we only keep the logarithmic divergences in the sum over $n$ --,\footnote{We thank R. Loganayagam and Shiraz Minwalla for this absolutely essential step.}
\begin{align}
 A(a,b) &= \sum_{j,r = 0}^{\infty} g_{j} (b) \left\{ (-a)^{r} + (a+b-P)^{r} \right\} \zeta (k(b) + j + r + 1) + reg. \nonumber\\
 &= \sum_{N=0}^{\infty} \sum_{J=0}^{N} g_{J} (b) \left\{ (-a)^{N-J} + (a+b-P)^{N-J} \right\} \zeta(k(b) + N + 1) + reg.
 \label{eqn:an-cont}
\end{align}
This expression now makes sense for $\Re b < 0$ as well.
While this analytic continuation can be justified merely on the grounds of being an analytic continuation and therefore unique, we also show in appendix \ref{sec:explicitVS} in more detail how it works.

The Riemann zeta function $\zeta(z)$ has a pole at $z = 1$ with residue $1$.
Thus, the expression \eqref{eqn:an-cont} has poles at $k (b) = -N$ with residues
\begin{equation}
  Res_{b = k^{-1} (-N)} A(a,b) = \frac{1}{k'(k^{-1}(-N))} \sum_{J=0}^{N} g_{J} (k^{-1} (-N)) \left\{ (-a)^{N-J} + (a+k^{-1}(-N)-P)^{N-J} \right\}.
  \label{eqn:residues}
\end{equation}

Consistency with crossing symmetry therefore gives us a condition on the function $k(b)$,
\begin{equation}
 k (-n) = -N, \text{so that there are poles at } b = -n,
 \label{eqn:k-fn-cond}
\end{equation}
and two countably infinite sets of conditions on the functions $g_{J}$,
\begin{align}
  k^{-1} (-N) \notin \{0,-1,\cdots\} &\!\! \Rightarrow \frac{1}{k'(k^{-1}(-N))} \sum_{J=0}^{N} g_{J} (k^{-1} (-N)) \left\{ (-a)^{N-J} + (a+k^{-1}(-N)-P)^{N-J} \right\} \nonumber\\[-5pt]
  & \hspace*{10cm}= 0 \nonumber\\
  k^{-1} (-N) = -n &\Rightarrow \frac{1}{k'(-n)} \sum_{J=0}^{N} g_{J} (-n) \left\{ (-a)^{N-J} + (a-n-P)^{N-J} \right\} = f_{n} (a).
  \label{eqn:gj-conds}
\end{align}
The first equation ensures that the spurious poles occuring at non-integer values of $b$ vanish, and the second ensures that the real poles have the correct residues, those required by channel duality.
Eqns \eqref{eqn:k-fn-cond} and \eqref{eqn:gj-conds} are the general conditions for duality.

\subsection{Discussion}
Before going ahead, we note that the linearity of the spectrum has not been substantially used anywhere yet.
In the case of a nonlinear spectrum, we can take $a$ to be a non-linear function of $s$ such that the poles are at $a = -n$ and similarly for $b$ and $c$.
The major difference in this case is that there's no expression of the form $c = P - a -b$; however, it is still true that $c$ can be determined given $a$ and $b$, since $a$ must still be an invertible function of $s$.
Then, eqns \eqref{eqn:k-fn-cond} and \eqref{eqn:gj-conds} will be valid with the modification that $a + k^{-1} (N) - P$ must be replaced by $c|_{b = k^{-1}(-N)}$.

Another point worth noting is that the function $k(b)$ has a physical meaning -- the residue $f_{n} (b)$ is a polynomial of degree $-k(-n)$, which means that $-k(-n)$ is the spin of the maximum spin particle exchanged at level $n$.
This might seem odd at first, since the function was defined in terms of not the physical variables $s,t,u$ but the made-up variables $a,b,c$.
The point is that the normalisation for $a$ -- that the poles are at $a=-n$ -- is important.
If we redefined $a$, then the degree of the $n^{th}$ polynomial wouldn't match the function in the exponent of the asymptotic behaviour.

Having noted that, we also note that these general conditions aren't very easy to solve.
Hence, we now restrict the asymptotic function $k(b)$ to be linear and proceed.

\section{The Case of a Linear Asymptotic Function} \label{sec:lin-asymptotics}
While we have found the general equations required for channel duality, eqns \eqref{eqn:k-fn-cond} and \eqref{eqn:gj-conds}, they aren't very easy to solve.
To facilitate control over the equations, we restrict to the case when the asymptotic function $k(b)$ is linear,
\begin{equation}
  k(b) = k b.
  \label{eqn:lin-asymp-fn}
\end{equation}
With this restriction, we show that the only consistent value of $k$ is $2$.
The main reason $k=2$ is special is that the duality equations simplify greatly at this value; we end this section with the simplest form of these equations.
The details for the case $k(b) = kb - l$ are not substantially different, so we drop it to aid clarity.

When $k(b)$ is a linear function, the conditon \eqref{eqn:k-fn-cond} is automatically satsified, and the countably infiite set of constraints \eqref{eqn:gj-conds} on the $g_{J}$s become
\begin{align}
  N \neq kn &\Rightarrow \sum_{J=0}^{N} g_{J} \left( - \frac{N}{k} \right) \left\{ (-a)^{N-J} + \left( a - \frac{N}{k} - P \right)^{N-J} \right\} = 0 \nonumber\\
  N = kn &\Rightarrow \frac{1}{k} \sum_{J=0}^{kn} g_{J} (-n) \left\{ (-a)^{kn-J} + (a-n-P)^{kn-J} \right\} = f_{n} (a).
  \label{eqn:gj-conds-lin}
\end{align}

We expand the residue $f_{n} (a)$ in powers of $-a$ as
\begin{equation}
  f_{n} (a) = \sum_{J = 0}^{kn} h_{J} (-n) (-a)^{kn-J}.
  \label{eqn:hj-defn}
\end{equation}
The choice of $-a$ instead of $a$ in this equation is parallel to the choice of expanding in $n$ instead of $-n$ in \eqref{eqn:1byn}.
Using this and the binomial expansion of $\left( a - \frac{N}{k} - P \right)^{N-J}$, we find
\begin{align}
 h_{J} \left( - \frac{N}{k} \right) &= \frac{1}{k} \left\{ g_{J} \left( - \frac{N}{k} \right) + (-1)^{N-J} \sum_{j=0}^{J} (-1)^{j} {N-J+j \choose j} \left( \frac{N}{k} + P \right)^{j} g_{J-j} \left( - \frac{N}{k} \right) \right\} \nonumber\\
 &\hspace{-1.4cm}= \frac{1}{k} \left\{ [1 + (-1)^{N-J}] g_{J} \left( - \frac{N}{k} \right) + (-1)^{N-J} \sum_{j=1}^{J} (-1)^{j} {N-J+j \choose j} \left( \frac{N}{k} + P \right)^{j} g_{J-j} \left( - \frac{N}{k} \right) \right\},
 \label{eqn:duality-eqns-coeffs}
\end{align}
with the definition
\begin{equation}
  h_{J} (\text{non-integer}) = 0.
  \label{eqn:hj-cheating}
\end{equation}
The first line in \eqref{eqn:duality-eqns-coeffs} is written in a form such that the $a$-channel and $c$-channel contibutions are separate from each other, while the second line collects the $g_{j}$s together.
We will henceforth use whichever form is convenient.

The reader may have noticed that the notation we've been using for the $h_{J}$s is (almost falsely) suggestive -- we've covertly treated them as functions of the level $n$.
The reason for this is that in the equations \eqref{eqn:duality-eqns-coeffs} the right hand side are in fact functions, and therefore we can promote $h$ to a function of the level by requiring equality with the right hand side.
This has to be consistent with the analytic continuation that allowed us to expand the residues $f_{n} (b)$ in a $1/n$ expansion.
But, notice that the right hand side has the factor $(-1)^{N}$, which has an essential singularity at $\infty$; this caontradicts our assumption about analyticity at $\infty$, as used in \eqref{eqn:1byn}.

For odd values of $k$, this poses an insurmountable problem.
Consider $k=1$ for example, so that the left hand side of \eqref{eqn:duality-eqns-coeffs} is non-zero for all values of $N$.
Then, the residue at level $n$ is
\begin{equation}
  f_{n} (a) = \sum_{J=0}^{n} g_{J}(-n) (-a)^{n-J} + (-1)^{n} \sum_{J=0}^{n} \sum_{j=0}^{J} (-1)^{j+J} {n-J+j \choose j} (n+P)^{j} g_{J-j} (-n) (-a)^{n-J}
  \label{eqn:k1}
\end{equation}
Because of the $(-1)^{n}$, this can't be analytic at $\infty$, assuming the $g_{j}$s are analytic at $\infty$.
However, we know that the $g_{j}$s are, in fact, analytic at $\infty$, which can be seen by taking the limit $b \to \infty$ of their definition \eqref{eqn:1byn}, where it must exhibit Regge behaviour.\footnote{One may object that \eqref{eqn:k1} is a polynomial not in $a$ but in $-a$; pulling out those signs merely shifts the problematic $(-1)^{n}$ to the other term and the non-analyticity remains.
It is also useful to remember to note that, while $\Gamma(z)$ has an essential singularity at $\infty$, the binomial coefficients diverge only polynomially.}
Thus, the channel duality equations for $k=1$ are inconsistent with our assumptions about analyticity at $\infty$.
It is easy enough to see that this extends to all odd values of $k$, and therefore that our assumptions aren't compatible with $k$ being odd.

What saves the case of even $k$ is that the physical poles are all at even values of $N$ in \eqref{eqn:duality-eqns-coeffs}; that means we can promote the $h_{J}$s to \emph{two} sets of functions, $h_{J}^{e}$ obtained from analytically continuing off even $N$ and $h_{J}^{o}$ obtained from analytically continuing off odd $N$; note here that the $h_{J}^{o}$s are identically $0$.

So we consider the equations for odd $N$ and even $N$ separately.
For odd $N$, the equations are
\begin{equation}
  \left\{ 1 - (-1)^{J} \right\} g_{J} \left( - \frac{N}{k} \right) = \sum_{j=1}^{J} (-1)^{J+j} {N-J+j \choose j} \left( \frac{N}{k} + P \right)^{j} g_{J-j} \left( - \frac{N}{k} \right), \quad N \text{ odd}.
  \label{eqn:odd-N}
\end{equation}
It is useful to again split these into two sets of equations, those in which $J$ is odd and those in which $J$ is even.
For even $J$, the left hand side is $0$ and we find
\begin{equation}
  0 = \sum_{j=1}^{J} (-1)^{j} {N-J+j \choose j} \left( \frac{N}{k} + P \right)^{j} g_{J-j} \left( - \frac{N}{k} \right),\quad N \text{ odd},\ J \text{ even}.
  \label{eqn:odd-N-even-J}
\end{equation}
And for odd $J$ we find
\begin{equation}
  g_{J} \left( - \frac{N}{k} \right) = - \frac{1}{2} \sum_{j=1}^{J} (-1)^{j} {N-J+j \choose j} \left( \frac{N}{k} + P \right)^{j} g_{J-j} \left( - \frac{N}{k} \right),\quad N \text{ odd},\ J \text{ odd}.
  \label{eqn:odd-N-odd-J}
\end{equation}
As can be seen, these are all constraints that relate various $g_{J}$s evaluated at the same point, for a countably infinite set of points.

As it happens, this is enough to argue that equations \eqref{eqn:odd-N-even-J} and \eqref{eqn:odd-N-odd-J} are valid everywhere -- that is, that they are functional relations among the $g_{J}$s.
This is possible because of a theorem in complex analysis called Carlson's theorem.
Roughly, it says that given a function specified at the positive integers, $f_{n}$, then there is a unique complex function $f(z)$ such that $|f(z)|$ grows at most as $e^{\tau |z|}$ for some $\tau$ at large $|z|$ with $\Re z > 0$ and $f(iy)$ grows slower than $e^{\pi |y|}$ at large values of $y$.

Given that the $g_{J}$s satisfy the conditions of this theorem, we can see that equations \eqref{eqn:odd-N-even-J} and \eqref{eqn:odd-N-odd-J} must be valid everywhere very easily.
Take, for example, eqn \eqref{eqn:odd-N-odd-J} and separately analytically both the left and the right hand sides.
The analytic continuation of the left hand side is clearrly $g_{J}$.
The analytic continuation of the right hand side is also unique, since it behave roughly as $\sum N^{2j} g_{J-j}$ and if the $g_{J}$s satisfy the conditions so do these.
Thus, eqn \eqref{eqn:odd-N-odd-J} is a functional relation.
One can similarly argue for eqn \eqref{eqn:odd-N-even-J}.

What remains is to show that the $g_{J}$s in fact satisfy the conditions of Carlson's theorem.
Since we are analytically continuing off the negative integers, we need to bound the growth of $g_{J} (b)$ in the left-half plane.
Suppose it grows exponentially on the left half plane, $g_{J} (b) \sim e^{- \nu b}$.
Then, because of the duality equations \eqref{eqn:duality-eqns-coeffs} and the definition \eqref{eqn:hj-defn} of the $h_{J}$s,
\begin{equation}
  f_{n} (a) \sim e^{\nu n} \sum_{J} (-a)^{kn-J}.
  \label{eqn:hypothetical-asymptotics}
\end{equation}
This blow-up at large $n$ contradicts Regge behaviour.
This means that the $g_{J}$s can't grow exponentially in the left-half plane, and by continuity can't grow expnonentially on the (upper or lower) imaginary axis either.
It is worth noting that this doesn't constrain the behaviour on the right-half plane; in particular, for the Virasoro-Shapiro amplitude, it behaves like $b^{b}$, which is super-exponential on the right-half plane and goes to $0$ on the left-half plane while being oscillatory on the imaginary axis.
Thus, we have proved that the relations \eqref{eqn:odd-N-odd-J} and \eqref{eqn:odd-N-even-J} are functional relations valid for all values of $N$ (but, remember, not $J$).

Having dealt with the odd $N$ equations, we can now deal with those for even $N$.
The major difference is that the $h_{J}^{e}$s that appear in these equations are not $0$.
The equations are
\begin{align}
 h_{J} \left( - \frac{N}{k} \right) &\!=\! \frac{1}{k} \left\{ \left[ 1 \!+\! (-1)^{J} \right] g_{J} \left( - \frac{N}{k} \right) \!+ (-1)^{J} \sum_{j=1}^{J}(-1)^{j} {N-J+j \choose j} \left( \frac{N}{k} + P \right)^{j} g_{J-j} \left( - \frac{N}{k} \right) \right\} \nonumber\\
 &= \frac{2}{k} g_{J} \left( - \frac{N}{k} \right),
 \label{eqn:even-N}
\end{align}
where we have used the functional relations \eqref{eqn:odd-N-even-J} and \eqref{eqn:odd-N-odd-J} -- for odd $J$ the first term in the first line vanished and the second term becomes the final answer, and for even $J$ the second term vanishes.

These are nearly the final forms of the duality equations.
To see where the final simplification comes from, consider the case $k=4$.
In this case, $N = 4n + 2$ are all spurious poles; thus, $h_{J} (-n - 1/2) = 0$ and therefore $g_{J} (-n-1/2) = 0$.
Since the $g_{J}$s are $0$ at an infinite number of evenly spaced points and $g_{J}$ can't grow exponentially, the only solution is $g_{J} (b) = 0$.
This argument clearly generalises to all values of $k$ except $2$, since for all $k > 2$ we can find an infinite set of evenly spaced points where $N$ is even and $N/k$ isn't an integer.

Thus, the final forms of the duality equations are
\begin{align}
  \text{Definitions: } &f_{n} (b) = \sum_{j=0}^{\infty} g_{j} (b) n^{-2b-j} = \sum_{J=0}^{2n} h_{J} (-n) (-b)^{2n-J}, \label{eqn:defns}\\
  \text{Residue-Matching Eqns: } &g_{j} (-n) = h_{j} (-n), \quad j \le 2n, \label{eqn:rmes}\\
  \text{Spurious-Pole Eqns: } &g_{J} (b) = - \frac{1}{2} \sum_{j=1}^{J} (-1)^{j} \frac{\Gamma(-2b -J + j + 1)}{\Gamma(j+1) \Gamma(-2b-J+1)} (P-b)^{j} g_{J-j} (b), \quad J \text{ odd}, \nonumber\\
  &0 = \sum_{j=1}^{J} (-1)^{j} \frac{\Gamma(-2b -J + j + 1)}{\Gamma(j+1) \Gamma(-2b-J+1)} (P-b)^{j} g_{J-j} (b),\quad\quad J \text{ even}. \label{eqn:spes}
\end{align}
The reason for naming the equations such is that, when $k=2$, all the equations from even $N$ involve matching the residues of poles that in fact exist in the amplitude and all those from odd $N$ are those that involve demanding that the residues of the spurious poles vanish.

These equations are among the main results of this note.

\subsection{Discussion}
The first order of business for this duiscussion section is to dispel an obvious objection, which is that the Veneziano amplitude
\begin{equation}
  B(a,b) = \frac{\Gamma(a) \Gamma(b)}{\Gamma(a+b)} \xrightarrow{a \to \infty} a^{-b},
  \label{eqn:false-veneziano}
\end{equation}
which flatly contradicts our assertion that the asymptotic behaviour has to be $a^{-2b}$.
The answer, of course, is that this is not actually the Veneziano amplitude; the full Veneziano amplitude is the sum of three terms
\begin{equation}
  A_{Ven} (a,b) = B(a,b) + B(a,c) + B(b,c),  \xrightarrow{a \to \infty} a^{-2b},
  \label{eqn:real-veneziano}
\end{equation}
In fact, the full Veneziano amplitude can be written in the functional form of the Virasoro-Shapiro amplitude, see \cite{Virasoro1:PhysRev.177.2309} for details.

Turning to other things, one might wonder if these duality equations \eqref{eqn:defns}, \eqref{eqn:rmes}, \eqref{eqn:spes} mean anything physically.
While it is hard to give an overarching narrative to these equations, various aspects of them reflect various physical facts.

Most obviously, the value of $k$ is related to the particle with highest spin at every level, as seen in \eqref{eqn:gj-conds-lin} for example -- the highest spin particle propagating at level $n$ has spin $kn$.
This throws some light on why $k=2$ is special.
If $k$ is odd, then levels at odd values of $n$ have a highest spin particle with odd spin, but odd spins can't propagate in the scattering of identical scalar bosons; thus, $k$ being odd would require an infinite number of intricate cancellations, and our analyticity arguments are essentially that they can't actually happen.
It's not as clear why $k=2$ is preferred over a generic even $k$.

First, note that the RMEs \eqref{eqn:rmes} can be obtained from the definitions of the $h_{J}$s and $g_{J}$s \eqref{eqn:defns} by a very simple prescription: interchanging $b$ and $-n$ interchanges the roles of the $h_{j}$s and $g_{j}$s in \eqref{eqn:defns} (up to problems with the summation limits).
It seems that a very simple prescription ensures channel duality.

The SPEs are equations that express the coefficients of odd powers of $b$ in terms of the coefficients of the even powers of $b$, as can clearly be seen by their explicit solutions\footnote{We show this numerically in an attached mathematica notebook.}
\begin{equation}
  g_{2j+1} (-n) = \sum_{p=0}^{j} (-1)^{p} (n+P)^{2p+1} \frac{2n+1-2j+2p}{2n-2j} Z(p+1) g_{2j-2p} (-n),  
  \label{eqn:spe-soln}
\end{equation}
where
\begin{equation}
  Z(x) = \frac{4}{\pi^{2x}} \sum_{k=-\infty}^{\infty} (4k+1)^{-2x} = \frac{4}{(4 \pi)^{2x}} \left\{ \zeta \left( 2x,\frac{1}{4} \right) + (-1)^{2x} \zeta \left( 2x, \frac{3}{4} \right) \right\}.
  \label{eqn:funny-z-thing}
\end{equation}
This fact suggests that the content of the SPEs is that there are no odd spin particles propagating in the amplitude.
And this is in fact the case; an arbitrary sum of even-spin Gegenbauer polynomials satsifies these equations, assuming the RMEs.

Thus, the value of $k$ encodes the spin spectrum of the exchanged paricles, the RMEs encode the actual non-trivial s-t crossing symmetry, and the SPEs (given the RMEs) encode the much simpler t-u crossing symmetry (which is equivalent to there not being any odd-spin particles).

\section{Solving the Bootstrap Equations} \label{sec:lol-max}
These equations can be solved for the scattering of four identical scalars.
In this section, we show that in this case there is an infinite-dimensional parameter space of solutions.
In particular, we show using the bootstrap equations that given a proposed amplitude $A_{0} (a,b,c)$ that is symmetric in its arguments and has the correct poles and asymptotics, any amplitude of the form
\begin{equation}
  A(a,b,c) = \sum_{m=0}^{\infty} a_{m} A_{m} (a,b,c) \equiv \sum_{m=0}^{\infty} a_{m} A_{0} (a+m,b+m,c+m)
  \label{eqn:most-general-apmplitude}
\end{equation}
also satisfies these same conditions \cite{Khuri:1970ax,Weimar:1974pi,Matsuda:PhysRev.185.1811,COON1969669}.

First, we note that this is clearly true.
The term $A_{m}$ has poles at $a=-m,-m-1,-m-2,\cdots$ and similarly for $b$ and $c$, all of which are the poles of $A_{0}$, and therefore the entire sum has the same set of poles as the first term.
Similarly, the $m^{th}$ term has asymptotic behaviour $n^{-2b-2m}$, which is dominated by the asymptotic behaviour of $A_{0}$.
Finally, it is manifestly crossing symmetric.
Thus, it satisfies all the bootstrap constraints apart from unitarity.
We now show this more explicitly using the bootstrap eqns \eqref{eqn:defns}-\eqref{eqn:spes}.

To usefully solve the bootstrap equations, we need to parametrise the residues in some fashion, so that we work with the minimum amount of independent data.
The natural way to parametrise residues in a scattering amplitude is as sums of Gegenbauer polynomials with different spins.
For our purpose, however, this is an inconvenient basis, because it is very hard to find the combination of Gegenbauers that has Regge asymptotic behaviour; even in the case of the simplest example of the Virasoro-Shapiro amplitude, the general decomposition of the residues into Gegenbauers is not known, and thus even in that case we can't ascertain exactly how the sum of Gegenbauers attains this behaviour.
Much more convenient would be a basis which has the correct asymptotic behaviour.
Luckily, there is one right at hand, that given by the $A_{m}$s in  \eqref{eqn:most-general-apmplitude}.
Using this decomposition, we'll see that there are no constraints on the coefficients $a_{m}$ coming from crossing symmetry.

We write the residue as
\begin{equation}
  f_{n} (b) = \sum_{m=0}^{n} a_{m} F_{n,m} (b), \quad \big/F_{n.m} (b) = F_{n-m,0} (b+m),
  \label{eqn:residue-decomposition}
\end{equation}
where $F_{n,m} (b)$ is the residue of $A_{m}$ at $a=-n$.
The expansion of $F_{n,m} (b)$ about $n=\infty$ is
\begin{equation}
  F_{n,m} (b) = \sum_{j=0}^{\infty} G_{j}^{m} (b) n^{-2b-j} \quad \big/ G_{j<2m}^{m} (b) = 0.
  \label{eqn:m-1byn}
\end{equation}
The condition on the $G_{j}^{m}$s comes from the fact that $A_{m} \sim n^{-2b-2m}$.

Because the $A_{m}$s satisfy the bootstrap equations themselves, we can apply the RMEs \eqref{eqn:rmes} to rerwrite the full residue as
\begin{align}
  f_{n} (b) &= \sum_{m=0}^{n} a_{m} F_{n,m} (b) \nonumber\\
  &= \sum_{m=0}^{n} a_{m} \sum_{j=0}^{2n} G_{j}^{m} (-n) (-b)^{2n-j} \nonumber\\
  &= \sum_{j=0}^{2n} \left\{ \sum_{m=0}^{n} a_{m} G_{j}^{m} (-n) \right\} (-b)^{2n-j}.
  \label{eqn:full-residue-detailed}
\end{align}
From here, we can again apply the RMEs to the full amplitude to find
\begin{equation}
  g_{j} (-n) = \sum_{m=0}^{n} a_{m} G_{j}^{m} (-n).
  \label{eqn:gj-param}
\end{equation}

Now,we can plug this form for the $g_{j}$ to find
\begin{align}
  \sum_{m=0}^{n} a_{m} G_{2J+1}^{m} (-n) &= \sum_{m=0}^{n} a_{m} \left\{ - \frac{1}{2} \sum_{j=1}^{2J+1} (-1)^{j} {2n-2J+j \choose j} (P+n)^{j} G_{2J+1-j}^{m} (-n) \right\} \nonumber\\
  0 &= \sum_{m=0}^{n} a_{m} \sum_{j=1}^{2J} (-1)^{j} {2n-2J+j+1 \choose j} (P+n)^{j} G_{2J-j}^{m} (-n).
  \label{eqn:final-spes}
\end{align}
We see that the expression multiplying each $a_{m}$ exactly vanishes by the fact that the $A_{m}$s satisfy crossing, and thus we've shown using the bootstrap equations that \eqref{eqn:most-general-apmplitude} is crossing symmetric.

\subsection{Discussion}
An important question the above analysis leaves unanswered is whether the form \eqref{eqn:most-general-apmplitude} is the most general allowed form of the amplitude.
For tachyons, it is, but not for particles of positive mass-squared.

One way we could have gone about solving the bootstrap equations might have been to parametrise the residues by an arbitrary sum of even-spin polynomials (polynomials that can be written as a sum of even-spin Gegenbauers).
Given that the bootstrap equations guarantee us that the residue at level $n$ is a polynomial of degree $2n$, and that the Gegenbauer with spin $s$ is a polynomial of degree $s$, the $n^{th}$ residue is a sum of $n+1$ even-spin Gegenbauers (of spins $0,2,\cdots 2n$); in other words, the residue at level $n$ is given by $n+1$ real numbers.

We already used such a paramterisation, \eqref{eqn:residue-decomposition}, in which the residue is given by $a_{m}, m=0,1\cdots n$.
Each polynomial used in this parametrisation, further, is guaranteed to be a positive sum of even-spin Gegenbauers, because of the $b-c$ ($t-u$) symmetry of the different components in \eqref{eqn:most-general-apmplitude}.
Thus, it naively seems that this is the most general parametrisation and thus \eqref{eqn:most-general-apmplitude} is the most general amplitude allowed by the bootstrap equations.

There is a problem, however, because some of these polynomials are in general either $0$ or a constant.
The problem stems because of the special kinematics of the three-point function of two particle at mass $m$ and one particle at mass $2m$.
By going to the rest frame of the heavy particle, one realises that both light particles have $0$ spatial momentum.
Because the incoming state has no orbital momentum, the heavy particle has to be a scalar.
Another way to see the same thing is that if the heavy particle weren't a scalar the interaction vertex would necessarily have some derivatives acting on the light particles, and because of the lack of momentum these derivatives owuld be $0$.
Further, in the case of three massless particles, the three-point function has to be $0$ on-shell.

Because of this, in the case of massless or massive particles, when the particle of mass $2m$ is necessarily in the spectrum, one of the residues of the original amplitude $A_{0}$ has a residue which isn't a polynomial of degree $2n$ but a constant.
And because of the structure of \eqref{eqn:residue-decomposition}, this means that every subsequent residue is short a polynomial.
Thus, it is not obviously true that \eqref{eqn:most-general-apmplitude} is the most general amplitude satisfying our conditions when the external particles aren't tachyons.
We have not been able to usefully add a polynomial at each level that corrects this problem to check whether it is the most general form, however.

\section{Unitarity} \label{sec:pictures}
Having shown that crossing symmetry allows for an infinite-dimensional parameter space of amplitudes, we may still hope that unitarity constrains it more.
While unitarity is hard to analyse in general, we numerically argue in this section that unitarity still alows for an open set in this infinite-dimensional paramter space -- that is, we argue that arbitrary small perturbations don't violate unitarity.
We can't make the argument in the general case, so we take the Virasoro-Shapiro amplitude in 4 spacetime dimensions and look at amplitudes of the form \eqref{eqn:most-general-apmplitude} with $A_{0}$ as the Virasoro-Shapiro amplitude.

First, we consider ampltiudes of the form $A_{0} + a_{m} A_{m}$ for a particular value of $m$.
The coupling constant of the $l^{th}$ Regger trajectory -- the set of particles of spin $2n-2l$ --, the coupling takes the form
\begin{equation}
  \lambda_{n,2n-2l}^{2} = f_{l} (n) + a_{m} g_{l} (n).
  \label{eqn:cpl-generic}
\end{equation}
The unitarity constraint, that this be positive, provides a lower bound for $a_{m}$ for values of $n$ such that $g_{l} (n)$ is positive and and upper bound when $g_{l} (n)$ is negative.
Maximising the lower bound and minimising the lower bound over all $n$ in the respective regions gives us the unitarity constraint on $a_{m}$ coming from the $l^{th}$ Regge trajectory.

We calculated it for $m=1,2\cdots 6$ using $8$ Regge trajectories.
The bounds, calculated in an attached mathematica notebook, are
\begin{align}
  -8.<&a_1<246\nonumber\\
  -1230.<&a_2<1744\nonumber\\
  -753408.<&a_3<545260\nonumber\\
  -6.19451\times10^8<&a_4<5.61988\times10^8\nonumber\\
  -1.04535\times10^{12}<&a_5<1.30753\times10^{12}\nonumber\\
  -3.5675\times10^{15}<&a_6<5.70538\times10^{15}.
  \label{eqn:unitarity-bounds-1}
\end{align}

\begin{figure}[t]
  \centering
    \includegraphics[width=60mm]{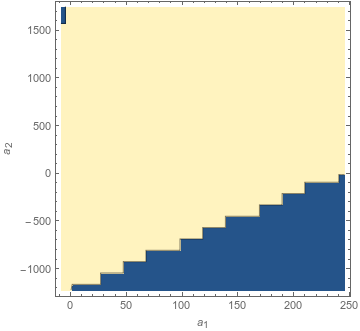}
    \includegraphics[width=60mm]{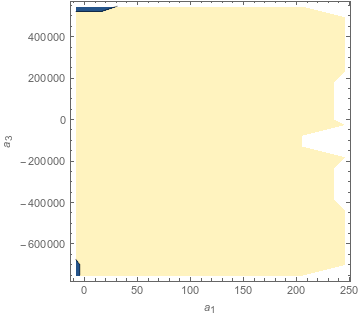}
    \includegraphics[width=60mm]{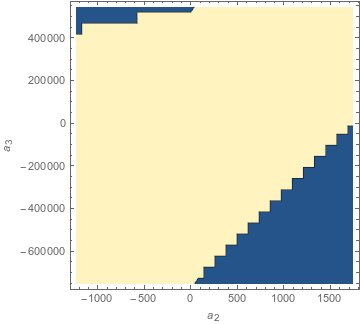}
  \caption{We checked unitarity for $25 \times 25$ points for amplitudes of the form $A_{0} + a_{m_{1}} A_{m_{1}} + a_{m_{2}} A_{m_{2}}$, using couplings from 5 Regge trajectories. The yellow regions are the ones allowed by unitarity, The three cases are (a) $m_{1}=1,m_{2}=2$ (b) $m_{1}=1,m_{2}=3$ (c) $m_{1}=2,m_{2}=3$. The ranges scanned were decided by the bounds shown in eqn \eqref{eqn:unitarity-bounds-1}.}
  \label{fig:unitarity-bounds-2}
\end{figure}

Second, we consider peturrbations with two subleading Virasoro-Shapriros at a time, $A_{0} + a_{m_{1}} A_{m_{1}} + a_{m_{2}} A_{m_{2}}$.
We take the values $(1,2)$, $(1,3)$ and $(2,3)$ for $(m_{1},m_{2})$.
Then, we consider $25\times 25$ points in the region allowed by eqn \eqref{eqn:unitarity-bounds-1} and check (again, a Regge trajectory at a time) whether each of these points is allowed by unitarity or not.
For this, we used $5$ Regge trajectories.
The allowed values of the parameters are coloured in yellow in fig \ref{fig:unitarity-bounds-2}.

\subsection{Discussion}
We see from the above that the unitarity of the Virasoro-Shapiro ampltude is stable to perturbations.
We did not run the checks for three or more subleading Virasoro-Shapiros because it seems pretty clear from the above numbers and graphs that unitarity doesn't constrain the amplitude very much.

One thing that we may note is that one reason the Virasoro-Shapiro amplitude is this stable to perturbations is that it doesn't seem to have any couplings that are $0$, \cite{natsuume1993natural}, and so there's always a perturbation small enough that it doesn't cause a problem.
Further, even if one of the couplings were $0$, it would only put a hard constraint on perturbation in one direction.
Thus, to get a non-trivial unitarity constraint, it must be that the same perturbation moves two couplings that are $0$ in oppposite directions.

\section{Conclusions} \label{sec:conclusions}
We have studied the conditions imposed by Regge asymptotic behaviour, crossing symmetry and unitarity on the scattering amplitude of four identical scalars in the case when the exchanged particles have a linear spectrum of exchanged particles.
We have found, in the general case, a countably infinite set of equations required for channel duality, \eqref{eqn:gj-conds}, on the residues of the amplitude.
Specialising to the case when the exponent in the Regge behaviour is a linear function of $b$, we have shown that the linear function must be $2b$, and further we've simplified the channel duality equations for this case to \eqref{eqn:defns}, \eqref{eqn:rmes} and \eqref{eqn:spes}.
We've shown that these equations allow an infinite-dimensional parameter space of solutions and that unitarity doesn't seem to impose strong enough constraints to help with the dimensionality of the parameter space.

It may well be the case, however, that this conclusion is too naive.
This is because in a theory one can calculate many different amplitudes featuring overlapping sets of cubic couplings.
To truly understand the conditions posed by unitarity, we must grapple with this whole morass of amplitudes.
Not only do we not know how to deal with the whole morass of amplitudes, we can't even perform the most basic step of explicitly proving the unitarity of the Virasoro-Shapiro amplitude without resorting to the worldsheet (in fact, the worldsheet was discovered in the process of investigating the unitarity of these amplitudes). Another interesting question would be to try and understand what happens if we demand that the spectrum of intermediate particles be only asymptotically linear. At present, we have no useful comments on this particular aspect and it is something that we would like to answer in a later work. Besides this, the original problem of understanding graviton scattering amplitudes still remains an important open problem.

We leave all these harder problems for future work.

\section{Acknowledgements}
We are grateful to S. Minwalla for suggesting this problem to us. We thank S. Jain, R. Loganayagam and S. Minwalla for being part of the original collaboration. We also thank Y-t Huang, A. Sen, A. Sinha and A. Zhiboedov for useful discussions. This work was presented as a poster in Strings 2017 and we thank the participants of the conference for stimulating discussions and feedback on this work. Authors are thankful to organisers of Indian Strings Meeting, 2016; Students Talks on Trending Topics in Theory, 2017; and, Strings 2017 for their hospitality when part of this work was in progress. This work was supported in part by Infosys Endowment for the study of the Quantum Structure of Space Time and by Indo-Israel grant of S. Minwalla which was graciously shared by him with us.

\appendix
\section{Kinematics} \label{sec:kinematics}
In this appendix, we collect some useful facts about the kinematics of the scattering we're considering.
The basic setup is the tree-level scattering of four identical scalars.
We ignore any four-point coupling between these scalars (since it's channel-dual already), and focus on the particles the scalars have a three-point coupling with.
These intermediate particles are exchanged in the $s,t$ and $u$ channels.
\subsection{Spinning intermediate particles}
\paragraph{Most general scalar-scalar-spin(\texorpdfstring{$l$}{l}) interaction}
The most general 3-point interaction between 2 scalars and a spin-$l$ particle is given by:
\def\del{\partial}
\def\lrdel{\overset{\leftrightarrow}{\partial}}
\def\ldel{\overset{\leftarrow}{\partial}}
\def\rdel{\overset{\rightarrow}{\partial}}

\begin{equation}\label{eq:gen-int}
	S_{int} = \lambda \int d^Dx \si^{\mu_1\cdots\mu_l}(x) \Big(\phi(x) \big( \lrdel_{(\mu_1}  \lrdel _{\mu_2} \ldots  \lrdel _{\mu_l)} \big) \phi(x) \Big), \quad \quad \lrdel =i( \ldel - \rdel )
\end{equation}
In writing the above interaction term we have taken into account the symmetric-transverse-traceless representation of an arbitrary spin-$l$ particle.\footnote{In symmetric-transverse-traceless representation, the polarization of the spinning particle can be expanded in a basis like:
 $\ep^{(\mu_1}\ep^{\mu_2}\cdots\ep^{\mu_l)}$, where $\ep\cdot\ep=0$, $\ep\cdot p=0$. Here $p$ is the momentum of the spinning particle and $\ep$ is its polarization.}

Note that the vertex identically vanishes when $l$ is odd. This happens because the above vertex picks up a sign $(-1)^l$ under the exchange of the two $\phi$ fields, which is basically a symmetry. Another way to see this is to consider a 3-point interaction as shown below. The amplitude should not change under the exchange of particle $1-2$. However, this corresponds to a rotation by an angle $\pi$ in the center of mass frame and the odd-spin particle picks up a phase, $(-1)$. Thus for consistency, this 3-point interaction vanishes identically.

\paragraph{Propagator of a spin-$l$ particle} Finding the propagator of a general spin-$l$ particle is a matter of projecting out a $l$-tensor in the correct symmetric-transverse-traceless representation and has been worked out in \cite{hayashi,Singh-hagen}. We quote the momentum space propagator here:
 \begin{eqnarray}
    \frac{-i\Theta^{(l)}_{\mu_1\dots\mu_s,\nu_1\dots\nu_l}\vert_{p^2=m^2}}{-p^2+m^2}
  \end{eqnarray}
  where $\Theta^{(l)}_{(\mu),(\nu)}$ is the  spin-$(l)$ analogue of the projection operator given by:
  \begin{eqnarray}\label{eq:prop-s}
    \Theta^{(s)}_{\mu_1\dots\mu_s,\nu_1\dots\nu_l}&=&\Bigg \{ \underset{p=0}{\overset{[l/2]}{\sum}}\frac{(-1)^p l!(2l + D -2p-5)!!}{2^pp!(l-p)!(2l + D -5)!!} 
    \Theta_{\mu_1\mu_2}\Theta_{\nu_1\nu_2}\dots\Theta_{\mu_{2p-1}\mu_{2p}}\Theta_{\nu_{2p-1}\nu_{2p}}\cr
    &&\hspace{1cm} \times \Theta_{\mu_{2p+1}\nu_{2p+1}}\dots \Theta_{\mu_l,\nu_l} \Bigg \}_{sym(\mu),sym(\nu)},
  \end{eqnarray}
  where $\Theta_{\mu\nu}=\eta_{\mu\nu}-p_\mu p_\nu/p^2$ is the spin one projection operator and $[l/2]$ gives the largest integer lesser than $l/2$. In the above expression $\{\cdot \}_{sym(\mu),sym(\nu)}$ denotes that the expression needs to be symmetrized in all the $\mu,\nu$ indices.
  
  \paragraph{4-scalar scattering with spin(\texorpdfstring{$l$}{l}) exchange} Using the expressions for the interaction between scalar and spin-$l$ in \eqref{eq:gen-int} and the propagator, \eqref{eq:prop-s}, one can easily write down the 4-scalar scattering amplitude with spin-$l$ intermediate particle.
   \begin{figure}[H]
  \centering
  \includegraphics[scale=0.8,trim=0cm 0.1cm 0cm 0cm,clip]{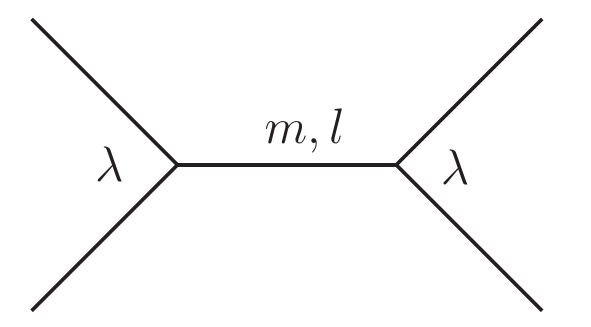}
  \caption{4-scalar scattering with an exchange of particle of spin ($l$) and mass ($m$)}
\end{figure}
\begin{equation}
\langle \phi(k_1) \phi(k_2) \phi(k_3) \phi(k_4) \rangle_l
= \lambda^2 \Bigg( \frac{l! \ \Gamma \pqty{ \frac{D-3}{2}} \pqty{s-4M^2}^l }{2^l \Gamma \pqty{ \frac{D-3}{2}+l}} \times C^{\frac{D-3}2}_l\bqty{ 1+\frac{2t}{s-4M^2} } \Bigg) \times \frac1{s-m^2}
\end{equation}
Here, $C^{\frac{D-3}2}_l$ are the Gegenbauer polynomials, which originate in the above expression due the particular structure of contractions that appears in \eqref{eq:prop-s}. The Gegenbauer polynomials obey the following orthogonality condition:
\begin{equation}
	\int\limits_0^\pi \!\! d\theta \qty(\sin\theta)^{D-3} \; C^{\frac{D-3}2}_l\bqty{ cos\theta } C^{\frac{D-3}2}_{l'}\bqty{ cos\theta } = 2^{4-D} \pi \frac{ \Ga\qty(l+D-3) }{ \qty(l+\frac{ D-3 }2) \Ga(l+1) \Ga\qty(\frac{D-3}2)^2 } \ \de_{ll'} 
\end{equation}
The residue of the pole at $s=m^2$ is given by,
\begin{equation}\label{eq:res-at-l}
	\lambda^2 \Bigg( \frac{l! \ \Gamma \pqty{ \frac{D-3}{2}} \pqty{m^2-4M^2}^l }{2^l \Gamma \pqty{ \frac{D-3}{2}+l}} \times C^{\frac{D-3}2}_l\bqty{ \cos(\theta) } \Bigg) 
\end{equation}
here,
$\cos(\theta)$ is the angle of scattering in the center of mass coordinates and is related to the Mandelstam variables by following relations, $$\cos(\theta) = 1 + \frac{2t}{s-4M^2} = \frac{u-t}{u+t}$$

\subsection{Threshold kinematics}
We show that for the scattering of massive scalar particles of mass $M$, there exists a threshold at mass $2M$, at which the kinematics becomes trivial. This is clear from the expression \eqref{eq:res-at-l} for the residue of a physical pole. If the pole occurs at $s = m^2 = 4M^2$, corresponding to a particle of mass $2M$, then the residue vanishes identically, for $l\neq0$. Thus, at such a threshold mass only scalar particles are allowed. Consequently, the residue, instead of being a polynomial of appropriate degree in $t$ or $\cos(\theta)$ is a constant. For the case of massless external particles the threshold particle is also massless and hence the residue vanishes identically even for $l=0$.

For the class of amplitudes that interest us: those one with a linear spectrum, the threshold condition is always met for particles with positive or zero mass. For such particles, the residue of the amplitude becomes a constant at some excited level. However, this doesn't happen for tachyonic particles, for which the threshold mass is a particle with an even more negative mass.

\section{Explicit Demonstration of Channel Duality} \label{sec:explicitVS}
In this appendix, we explicitly show how channel duality works in the case of the Euler Beta function
\begin{equation}
  B(a,b) = \frac{\Gamma(a) \Gamma(b)}{\Gamma(a+b)},
  \label{eqn:beta-fn}
\end{equation}
which is the building block of the Veneziano amplitude.
This serves both as an intuition-building exercise, and as a demonstration of the vailidity of our techniques in the main text.
We choose this function for simplicity; it doesn't have any $c$-poles, which makes the equations significantly shorter.
However, precisely because it doesn't have these poles, it also lacks full crossing symmetry under arbitrary permutations of $a,b,c$; it is, however, still invariant under the exchange of $a$ and $b$.

First, we go to the region $a<0, b>0$, which is the physical s-channel scattering regime.
There, we may write
\begin{equation}
  B(a,b) = \sum_{n=0}^{\infty} \frac{1}{a+n} \frac{(1-b)(2-b)\cdots (n-b)}{n!}.
  \label{eqn:ven-expanded}
\end{equation}
This sum converges for $\Re b > 0$, since the summand behaves for large $n$ as $n^{-b-1}$.

At the edge of this region of convergence, $b=0$, all the residues in the expansion are $1$.
We can recreate this result from the $t$-channel expansion, valid for $a>0,b<0$,
\begin{equation}
  B(a,b) = \sum_{n=0}^{\infty} \frac{1}{b+n} \frac{(1-a)(2-a)\cdots (n-a)}{n!}
  \label{eqn:ven-expanded-b}
\end{equation}
by the following (strictly invalid) trick.
First, we take $b=0$ and $a=-m$ (which, notice, is outside the region of convergence); the expression \eqref{eqn:ven-expanded-b} then becomes
\begin{align}
  B(a,0) & \xrightarrow{a \to 0} \sum_{n} \frac{1}{n} \nonumber\\
  & \xrightarrow{a \to -1} \sum_{n} 1 + \frac{1}{n}  \nonumber\\
  & \xrightarrow{a \to -2} \sum_{n} \frac{n}{2} + \frac{3}{2} + \frac{1}{n} \nonumber\\
  & \xrightarrow{a \to -m} \sum_{n} \cdots + \frac{1}{n}.
  \label{eqn:fun-recreation}
\end{align}
It seems that the coefficient of the $\sum \frac{1}{n}$ term gives the correct residue for this value of $b$.
While recreating a factor of $1$ is neither useful nor kosher, it is still valid for the reason that we can recreate it using a lot less arbitrary prescription: replacing $\sum n^{-s}$ by the Riemann zeta function $\zeta(s)$, which has a simple pole at $s=-1$!
This prescription is less arbitrary for the simple reason that it provides an analytic continuation off the right-half-plane, and it is therefore the unique prescription.

It is useful to do this in a more systematic manner.
We rewrite the $s$-channel expansion as
\begin{equation}
  B(a,b) = \sum_{n=0}^{\infty} \frac{f_{n} (b)}{a+n} \quad \Big/ f_{n} (b) = \frac{\Gamma(n+1-b)}{\Gamma(n+1) \Gamma(1-b)}.
  \label{eqn:ven-binomial-expansion}
\end{equation}
We can expand the residues $f_{n} (b)$ in a $1/n$-series,
\begin{equation}
  f_{n} (b) = \sum_{j=0}^{\infty} g_{j} (b) n^{-b-j}, 
  \label{eqn:1byn-ven}
\end{equation}
where, for reference, the first few $g_{j}$s are
\begin{align}
  g_{0} (b) &= \frac{1}{\Gamma(1-b)} \nonumber\\
  g_{1} (b) &= \frac{b(b-1)}{2\Gamma(1-b)} \nonumber\\
  g_{2} (b) &= \frac{b (2 - 3 b - 2 b^2 + 3 b^3)}{24 \Gamma(1-b)} \nonumber\\
  g_{3} (b) &= \frac{(-1 + b)^2 b^2 (2 + 3 b + b^2)}{48 \Gamma(1-b)} \nonumber\\
  g_{4} (b) &= \frac{b (-48 + 20 b + 180 b^2 - 25 b^3 - 192 b^4 - 10 b^5 + 60 b^6 + 
    15 b^7)}{5760 \Gamma(1-b)}.
  \label{eqn:ven-gjs}
\end{align}
The manipulations from eqn \eqref{eqn:1byn-expanded-1} to eqn \eqref{eqn:gj-conds} go through essentially as in the main text, except without the terms coming from the poles in $c$, with $k(b) = b$.
Eqn. \eqref{eqn:gj-conds} then becomes
\begin{equation}
  f_{n} (b) = \sum_{j=0}^{n} g_{j} (-n) (-b)^{n-j},
  \label{eqn:ven-final-answer}
\end{equation}
which are exactly the residue-matching equations \eqref{eqn:rmes}.
As an aside, we note the lack of any spurious-pole equations \eqref{eqn:spes}; this is because of the lack of poles in $c$, giving further credence to their interpretation as arising from $b-c$ symmetry.

The reader may readily check using the $g_{j}$s listed here that eqn \eqref{eqn:ven-final-answer} is indeed correct.
A much simpler way to check is to note the eqn \eqref{eqn:ven-final-answer} is the same as eqn \eqref{eqn:1byn-ven} with $b$ and $-n$ interchanged.
The exact expression for the residue, eqn \eqref{eqn:ven-binomial-expansion}, is manifestly invariant under that interchange, and therefore we must have arrived at the right answer.

\bibliographystyle{JHEP} 
\bibliography{paper1} 
\end{document}